\documentclass[a4paper,alpha-refs]{aejstyles}

\usepackage{amsmath}
\usepackage{graphicx}
\usepackage{siunitx}
\usepackage{xcolor, soul}

\newcommand{\shortcomment}[3]{\textcolor{#1}{[#2: #3]}}
\newcommand{\cefcom}[1]{\shortcomment{red}{CEF}{#1}}     

\sethlcolor{yellow}

\newcommand{\apjl}{\textit{ApJ}}
\newcommand{\apjs}{\textit{ApJS}}

\journal{aej}

\title{\texttt{MESA-Web}: A cloud resource for stellar evolution in astronomy curricula}

\author[1,2\authfn{1},\authfn{2}]{Carl~E. ~Fields}
\author[3]{Richard H. D. Townsend}
\author[4]{A.L.~Dotter}
\author[5]{Michael Zingale}
\author[6]{F.X.~Timmes}

\affil[1]{Computer,  Computational,  and Statistical Sciences Division,  Los Alamos National Laboratory,  USA}
\affil[2]{Steward Observatory,  University of Arizona,  USA}
\affil[3]{Department of Astronomy,  University of Wisconsin-Madison,  USA}
\affil[4]{Department of Physics and Astronomy, Dartmouth College, USA}
\affil[5]{Department of Physics and Astronomy, Stony Brook University, Stony Brook, USA}
\affil[6]{School of Earth and Space Exploration,  Arizona State University, USA}

\authnote{\authfn{1}Feynman Fellow}
\authnote{\authfn{2}carlnotsagan@lanl.gov}

\papercat{Resources \& Activities}

\runningauthor{Fields et al.}

\jvolume{00}
\jnumber{0}
\jyear{2023}

\begin{document}

\begin{frontmatter}
\maketitle

\begin{abstract}
We present \href{http://user.astro.wisc.edu/~townsend/static.php?ref=mesa-web-submit}{\texttt{MESA-Web}}, 
a cloud resource with an online interface to the Modules for Experiments in Stellar Astrophysics (\href{https://docs.mesastar.org/en/release-r22.05.1/}{\texttt{MESA}}) software instrument. 
\texttt{MESA-Web} allows learners to evolve stellar models without the need to download and install \texttt{MESA}. Since being released in 2015, 
\texttt{MESA-Web} has delivered over 17,000 calculations to over 2,200 unique learners and currently performs about 11 jobs per day.
\texttt{MESA-Web} can be used as an educational tool for stars in the classroom or for scientific investigations.
We report on new capabilities of \texttt{MESA-Web} introduced since its 2015 release including
learner-supplied nuclear reaction rates, custom stopping conditions, and an expanded selection of input parameters. 
To foster collaboration we have created a Zenodo  \href{https://zenodo.org/communities/mesa-web/}{ \texttt{MESA-Web} community hub} 
where instructors can openly share examples of using \texttt{MESA-Web} in the classroom.
We discuss two examples in the current community hub. The first example is a lesson module on Red Giant Branch stars 
that includes a suite of exercises designed to fit a range of learners and a \texttt{Jupyter} workbook for additional analysis.
The second example is lesson materials for an upper-level Astronomy majors course in Stars and Radiation
that includes an assignment verifying some of the expected trends that are presented in a popular stellar physics textbook.
\end{abstract}


\end{frontmatter}


\section{Introduction} \label{sec:intro}
One of the cornerstones underpinning modern astrophysics is the
fundamental properties of stars throughout their evolution.
Transformative capabilities in space- and ground-based hardware
instruments are providing an unprecedented volume of high-quality
measurements of stars, significantly strengthening and extending the
observational data upon which stellar astrophysics ultimately depends.
Revolutionary advances in software infrastructure, computer processing power,
and data storage capability are enabling
exploration of gravitational waves from the mergers of neutron stars and black holes,
revolutionary new sky surveys that probe ever-larger areas of the dynamic sky and ever-fainter sources,
the oscillation modes of stars across the Hertzsprung-Russel (HR) diagram, and more.

The standard computational tool of anyone interested in understanding stars is a stellar evolution code - a piece of software that can construct a model for the interior of a star, and then evolve it over time. Evolution codes allow us to check and refine the various physical theories that together compose stellar astrophysics (e.g., atomic physics, nuclear physics, fluid dynamics, thermodynamics); they provide laboratories for performing experiments on stars (e..g, discovering what factors contribute to the formation of red giants); and, they shed light on stages of stellar evolution that may be too fleeting to observe directly in the Universe.

While a number of stellar evolution tools exist, many are proprietary and unavailable outside restricted research settings. 
The past decade has seen the rise of sophisticated, open-source (anyone can freely download),
open-knowledge (best practices are freely shared), community-driven software instruments that model 
stars throughout their evolution 
\citep[e.g., Modules for Experiments in Stellar Evolution, \texttt{MESA}, ][]{paxton_2011_aa,paxton_2013_aa,paxton_2015_aa,paxton_2018_aa,paxton_2019_aa,jermyn_2023_aa}.
Moreover, the \texttt{MESA} Project supports a vibrant global community of learners and researchers (1000+ registered users), 
with active mailing lists, distributed version control, summer schools, and other learning activities.

Ubiquitous usage of a stellar evolution instrument in
classrooms worldwide remains a rich site of fascinating challenges.
For example, stellar evolution software instruments can be complicated
to install.  They require comfort using the Unix command line, setting
up compilers and other required software elements, and installing the
software instrument.  The hardware must also be capable enough 
to calculate the evolution in a reasonable amount of wall-clock time.  These can be barriers, 
especially in under-resourced communities and/or when a course
objective is aimed at a pedagogical survey of the evolution of stars.
Our goal is to lower these barriers to entry and to provide open educational resources
for learning about stellar evolution.

\texttt{MESA-Web}\footnote{\href{http://user.astro.wisc.edu/~townsend/static.php?ref=mesa-web-submit}{http://user.astro.wisc.edu/~townsend/static.php?ref=mesa-web-submit}}
 is a cloud resource featuring a point-and-click web interface
that lowers the barrier to entry of using \texttt{MESA} for education.
\texttt{MESA-Web} supports a variety of options for evolving a stellar model.
Upon completion, \texttt{MESA-Web} delivers to the learner an MP4 movie 
containing numerous model diagnostics and plain text output files for use in many common plotting package.
Through hands-on exercises, learners can gain new knowledge about the 
properties of stars that are not possible with traditional textbook-only approaches.
\texttt{MESA-Web} offers unique opportunities to dynamically illustrate 
the evolution of stars by following a golden rule of learning: involve don’t tell.

This article is organized as follows.  
In \S~\ref{sec:evolution}, we discuss major improvements and milestones since the creation of  \texttt{MESA-Web}.
In \S~\ref{sec:implementation} we discuss aspects of our \texttt{MESA-Web} implementation for
educators who may want to adopt it for their own software instruments.
In \S~\ref{sec:capabilities} we discuss the current capabilities of \texttt{MESA-Web},
in \S~\ref{sec:output} we discuss working with the stellar model output, and 
in  \S~\ref{sec:classroom} we highlight learner exercises and describe resources
for instructors to share their \texttt{MESA-Web} exercises.

\section{Origins to the Present}
\label{sec:evolution}

\begin{figure}[!htb]
\centering
\includegraphics[width=\columnwidth]{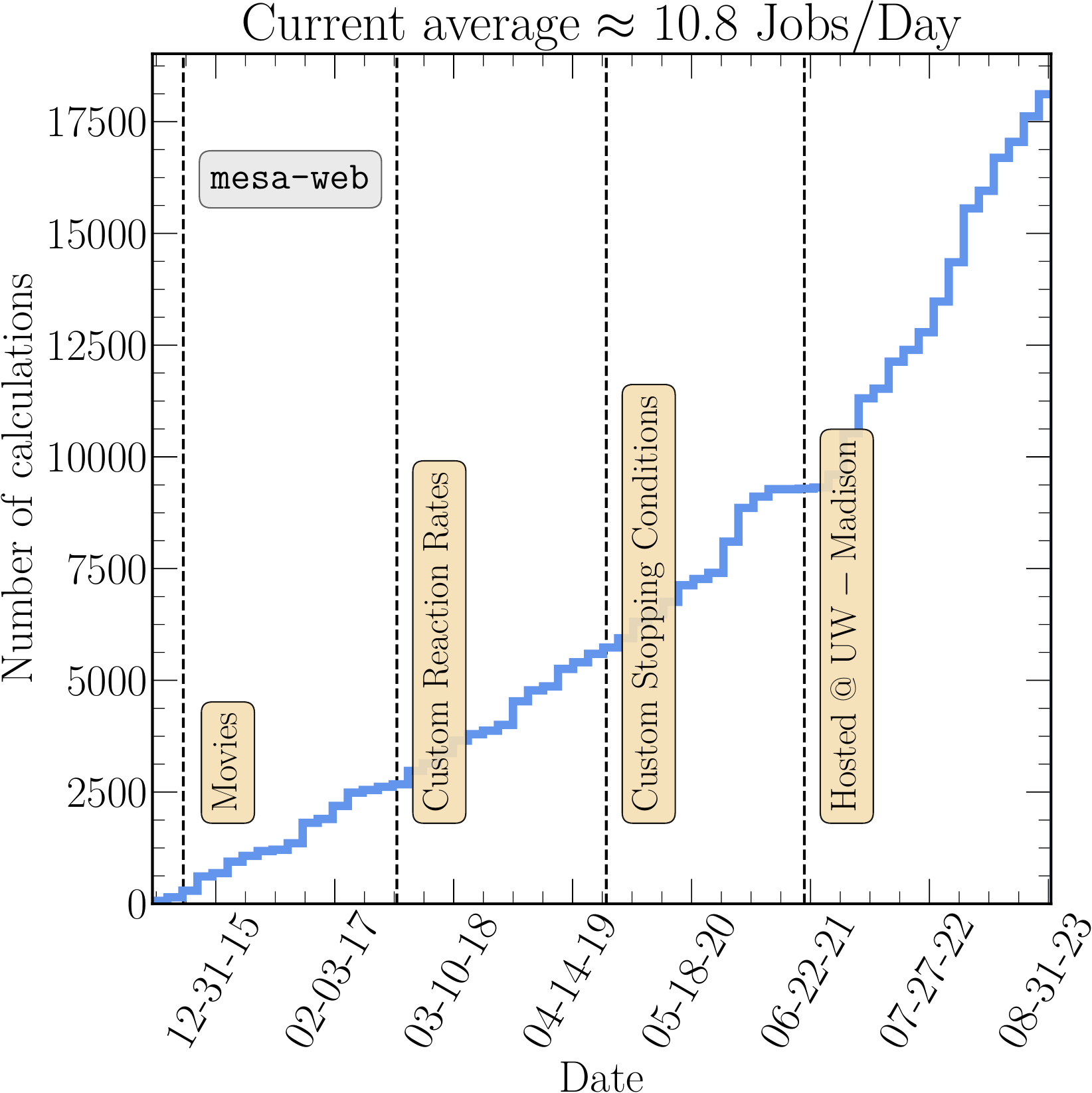}
\caption{
\texttt{MESA-Web} usage to date.
Dashed vertical lines and labels indicate the dates when a new capability was released.
}
\label{fig:growth}
\end{figure}

\texttt{MESA-Web} launched in June 2015 on a 2 core, 4 GB RAM, 10 GB
disk server hosted by Arizona State University.  Responding to
community demand for increased throughput, the platform was upgraded
in 2017 for a modest additional cost to a 4 core, 8 GB RAM, 50 GB disk
configuration.  Still under-powered relative to average community
demand, in August 2021 \texttt{MESA-Web} was re-factored and ported to
a 48 core, 192 GB RAM, 2 TB disk cluster at the University of
Wisconsin-Madison. This new cluster supports 12 concurrent jobs, meets
current peak demand during traditional semesters, and substantially
enhances the computational resources focused on education.

Figure \ref{fig:growth} shows the cumulative growth of \texttt{MESA-Web} usage over a 8 year period.
To date, \texttt{MESA-Web} has provided more than 17,000 calculations to an international community 
of learners in courses at more than 50 institutions worldwide.
The University of Chicago, Stony Brook University, the University of California, Santa Cruz, 
University of Pittsburgh, Texas A\&M University Commerce are among the institutions that have 
most utilized \texttt{MESA-Web} since its first release. 
These metrics suggest that \texttt{MESA-Web} is having a growing impact in astronomy education.

Figure \ref{fig:growth} shows an MP4 movie file included with the output was introduced early in the evolution of  \texttt{MESA-Web}.
Following this development, we introduced examples of the Test Suite that come included with a standard \texttt{MESA} distribution. 
Because of the complex nature of some of the Test Suites (multiple inlists, high computational demand, or others) 
pre-computing the examples provided is the most efficient means of delivering these data to learners.

\begin{figure}[!htb]
\centering
\includegraphics[width=\columnwidth]{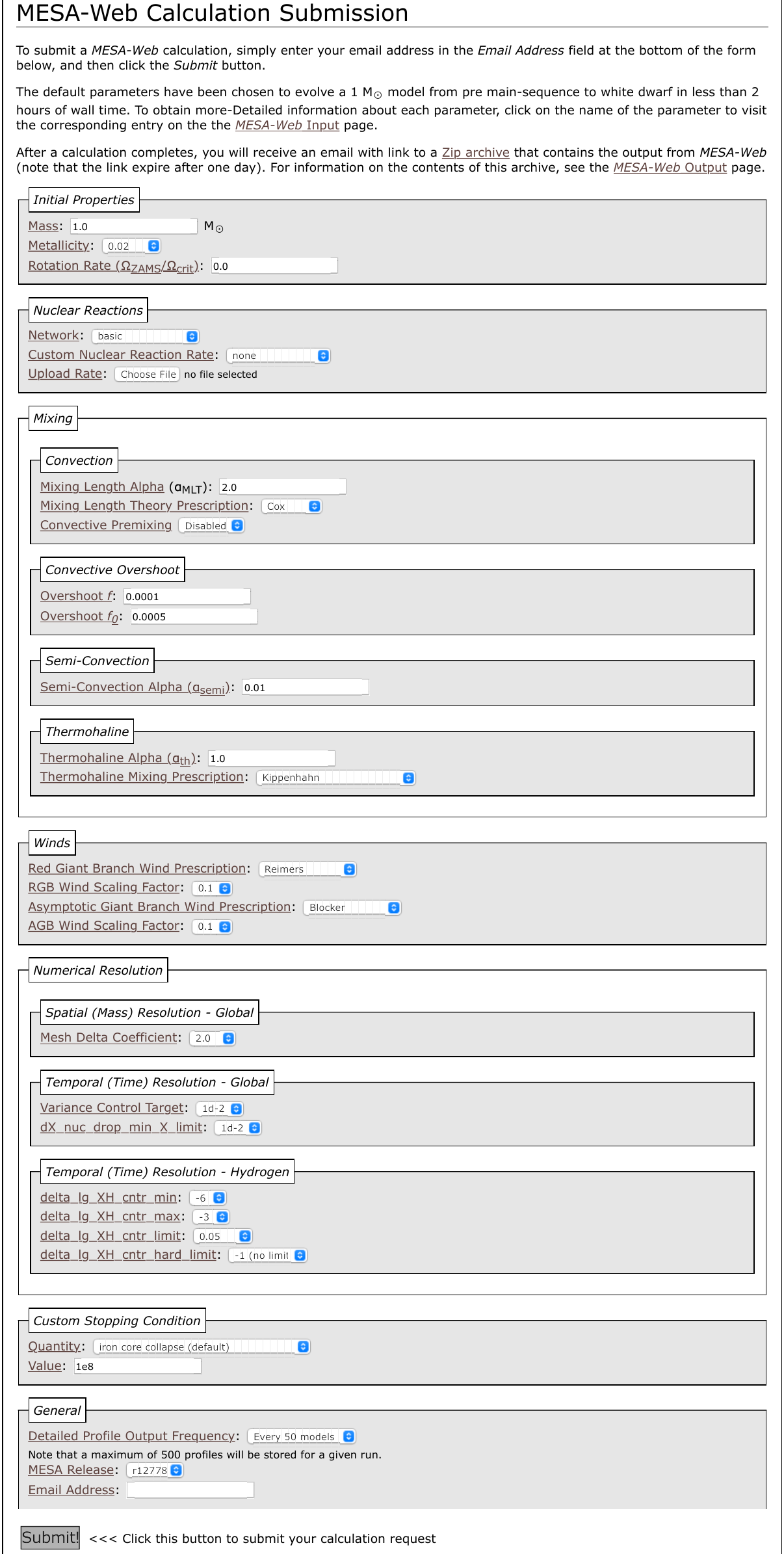}
\caption{
\href{http://user.astro.wisc.edu/~townsend/static.php?ref=mesa-web-submit}{\texttt{MESA-Web}} submission page.
}
\label{fig:submission}
\end{figure}

Figure \ref{fig:growth} also shows addition in mid-2017 of the option for a learner to choose a specific nuclear reaction from a list of eight key nuclear reactions and provide their own custom reaction rate.  This option is also visible in the submission form shown in Figure~\ref{fig:submission}.
The eight key reactions 
are the H-burning CNO cycle reactions
$^{12}$C($p$,$\gamma$)$^{13}$N, 
$^{13}$C($p$,$\gamma$)$^{14}$N,
$^{14}$N($p$,$\gamma$)$^{15}$O ;
the He-burning reactions
$\alpha$($\alpha$$\alpha$,$\gamma$)$^{12}$C ,
$^{12}$C($\alpha$,$\gamma$)$^{16}$O ,
$^{16}$O($\alpha$,$\gamma$)$^{20}$Ne ;
and the C-burning reactions
$^{12}$C($^{12}$C,$\alpha$)$^{12}$Ne, and 
$^{24}$Mg($\alpha$,$^{12}$C)$^{16}$O.
These eight are chosen as they have demonstrated significant impact on stellar properties
across a range of stellar environments including massive stars \citep{deboer_2017_aa,fields_2018_aa} 
and stars that produce white dwarfs \citep{luna_2006_aa,Fields_2016}.
If this option is chosen, a learner uploads a file containing the customized reaction rate data. This file must have 
(a) zero or more lines beginning with a hash (\#) mark; these are treated as comments and ignored;
(b) exactly one line consisting of a single integer, giving the number of subsequent lines.
(c) one or more lines containing a space-separated T8/rate pair, where T8 is the temperature in units of 10$^8$~K
and rate is the reaction rate in units of per second. Interpolation is used to evaluate the rate between the tabulated values.
A learner may use experimental or theory vales for the rates, including zero.

This capability also expanded \texttt{MESA-Web}'s impact to being a viable means of scientific investigation for the
nuclear astrophysics community without significant prior experience with \texttt{MESA}.

Finally, Figure \ref{fig:growth} shows the impact of providing a custom stopping condition in mid-2019. 
This option is also visible in the submission form shown in Figure~\ref{fig:submission}.
Learners can specify a limit to different stellar quantities as condition for which the model will complete. 
This new capability provides learners an increase in throughput especially if investigating a particular evolutionary epoch.  
\texttt{MESA-Web} can currently use  \texttt{MESA} versions 11701 or 12115 and is continually
updated to reflect a modern version.

\section{Technical Implementation}
\label{sec:implementation}

To provide an overview of the technical implementation of \texttt{MESA-Web},
here we narrate the series of steps involved in a typical calculation. The interface to \texttt{MESA-Web} is presented as web form
(Fig. 2), in which a learner can enter calculation parameters (e.g., initial mass, composition, stopping condition, nuclear reaction
network). A  \href{http://user.astro.wisc.edu/~townsend/static.php?ref=mesa-web-input}{link} to explanatory text
is provided for each parameter, indicating semantics, units and acceptable values. Sensible default
values are provided for all parameters save for the email address to
which results will be sent; taken together, these defaults result in a
simulation of the Sun's evolution.

When a learner submits the web form, a PHP script transmits the form data
from the web server to a separate computation server, using a TCP
socket. (The separation of web and calculation servers is both for
security and for computational efficiency). The computation server
validates the form data, checking each parameter lies within
acceptable bounds. If the data are valid, then it schedules a
calculation request and transmits a confirmation message back to the
web server; otherwise, it transmits an error message back that
indicates which parameters are invalid.

Calculation requests are fulfilled by a cluster comprising the
computation server (2 Intel Xeon E-2660 8-core CPUs, 64 GB RAM) and
two identical peer nodes, together with shared disk storage. The
cluster is managed using \texttt{SLURM} \citep{Yoo:2003}, an open-source job
scheduling system. Each calculation request is comprised of a pair of jobs:

\begin{description}
  \item[\emph{star}] --- running \texttt{MESA} with the supplied parameters (assigned 2 CPUs
    and a maximum runtime of 2 hours).
  \item[\emph{proc}] --- post-processing the \texttt{MESA} output (assigned 1 CPU) and emailing the learner.
\end{description}

Completion of the \emph{star} job is a pre-requisite for the
corresponding \emph{proc} job to run. The resources (CPU and time
limit) assigned to \emph{star} jobs is chosen to strike a balance
between how far a single calculation request is allowed to proceed,
and how many separate calculation requests can be completed within a
given time period; with these resources, \texttt{MESA} can follow the
evolution of a solar model from the pre-main sequence to the tip of
the asymptotic giant branch. 

The \emph{proc} job assembles the plot images written out by
\texttt{MESA} into an MP4 movie, and then packages this movie into a Zip
archive. This archive also includes a \texttt{MESA} `history' file, which
tabulates variables (e.g., surface radius and luminosity; central
elemental abundances) as a function of stellar age; and a series of
`profile' files, which tabulate variables (e.g., pressure,
temperature, density) as a function of position at selected instants
during the star's evolution. To facilitate analysis of these files, we
provide a Python module \texttt{mesaweb.py} on the
\texttt{MESA-Web} site. This module includes routines that read profile and
history data into Python dict data structures.

On completion of the \emph{proc} job, the Zip archive is copied back
to the web server, and an email is sent to the learner providing a
download link. This link remains valid for 24 hours, after which the
archive is deleted to free up space on the server.

As the functionality of the underlying \texttt{MESA} code grows and
changes, it is desirable to expose these improvements in
\texttt{MESA-Web}. However, this has to be balanced against the pedagogical
need to offer a stable and predictable service; a calculation request
re-submitted to \texttt{MESA-Web}, with the same parameters as before, should
provide exactly the same results no matter how much time has elapsed
between the two submissions. To strike this balance, \texttt{MESA-Web} offers
the ability to dynamically select which release of \texttt{MESA} is used
to service each calculation request.

\begin{figure*}[!htb]
\centering
\includegraphics[width=2.0\columnwidth]{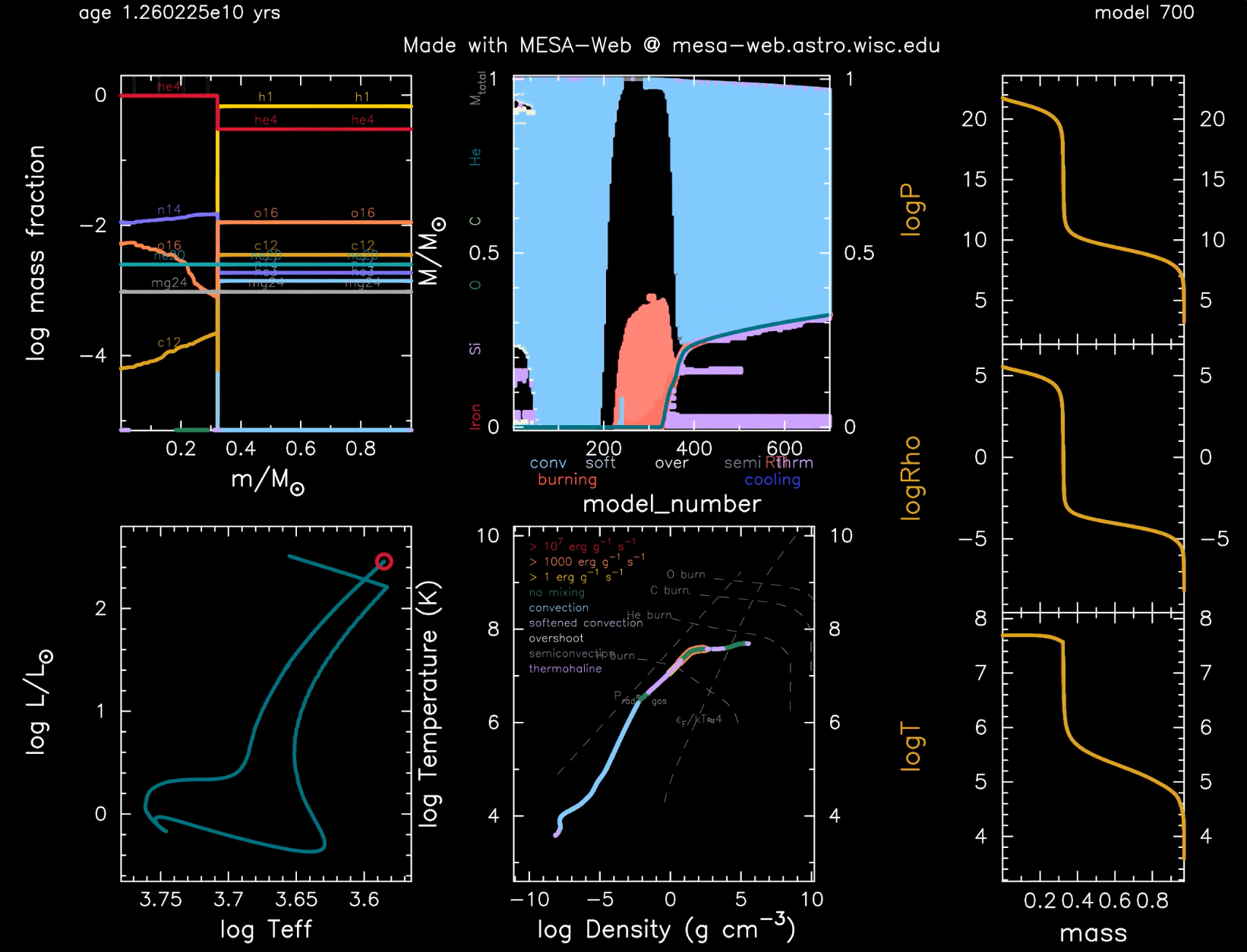}
        \caption[Grid plot of a \texttt{MESA-Web} calculation]{
Snapshot showing grid plot of profile and history data for \texttt{MESA-Web} calculation of a 1 $M_{\odot}$ rotating stellar model. At this model number, the star is on the red giant branch. The data shown in the movie are divided into five panels:
{\it upper-left} -- abundance profiles, plotting the mass fractions of selected nuclides as a function of mass coordinate within the star;
{\it upper-mid} -- a Kippenhahn (convective and burning profile history) diagram;
{\it lower-left} -- a Hertzsprung-Russell Diagram;
{\it lower-mid} -- a density-temperature profile plot;
{\it right} -- thermodynamic state profiles, plotting the total pressure (upper), density (mid), and temperature (lower) as a function of mass coordinate within the star.
}
\label{fig:grid}
\end{figure*}

\section{Learning with \texttt{MESA-Web} Capabilities}
\label{sec:capabilities}
Figure~\ref{fig:submission} shows the \texttt{MESA-Web} submission webpage as of  \today. 
Learners can specify different physical parameters of the stellar model including initial mass, 
metallicity, and mixing values.  Learners can also change the mass or temporal resolution,  
stopping conditions, and frequency of the returned data output. 
With these options learners can target specific stellar phenomena within the available computing resources.


The default values in the calculation submission page will evolve a 1 $M_{\odot}$ stellar model from the pre main-sequence to the tip of
the asymptotic giant branch within the walltime limit.   
For other unique evolutionary epochs, learners can consult the \href{http://user.astro.wisc.edu/~townsend/static.php?ref=mesa-web-input}{\texttt{MESA-Web} Input} page
where they can experiment with the input parameters that help build the stellar model and some details of their implementation in \texttt{MESA}.

Some advanced evolutionary epochs may require the use of a larger nuclear reaction network. Learners can
explore optional choices for a reaction network using the guidance at \href{http://user.astro.wisc.edu/~townsend/static.php?ref=mesa-web-nets}{\texttt{MESA-Web Nets}}. 
Larger networks take walltime to take a timestep and could potentially 
require reduced mass resolution to reach the phenomena within walltime limits. Advanced learners can consult the \texttt{MESA} instrument
papers for guidance in these choices \citep{paxton_2011_aa,paxton_2013_aa,paxton_2015_aa,paxton_2018_aa,paxton_2019_aa,jermyn_2023_aa}.


\texttt{MESA-Web} reduces barriers to accessing the core capabilities of \texttt{MESA} for any community.
One example is serving as a lightweight tool for researchers to produce stellar tracks to model the 
observational data of the eclipsing binary IT Librae \citep{wysocki_2022_aa}.  
Another example is enabling the experimental low-energy nuclear astrophysics community with a key scientific tool by providing the 
capability to utilize custom nuclear reaction rates for a list of key nuclear reactions common to many stellar phenomena.

\begin{figure*}[!t]
\centering
\includegraphics[width=2.0\columnwidth]{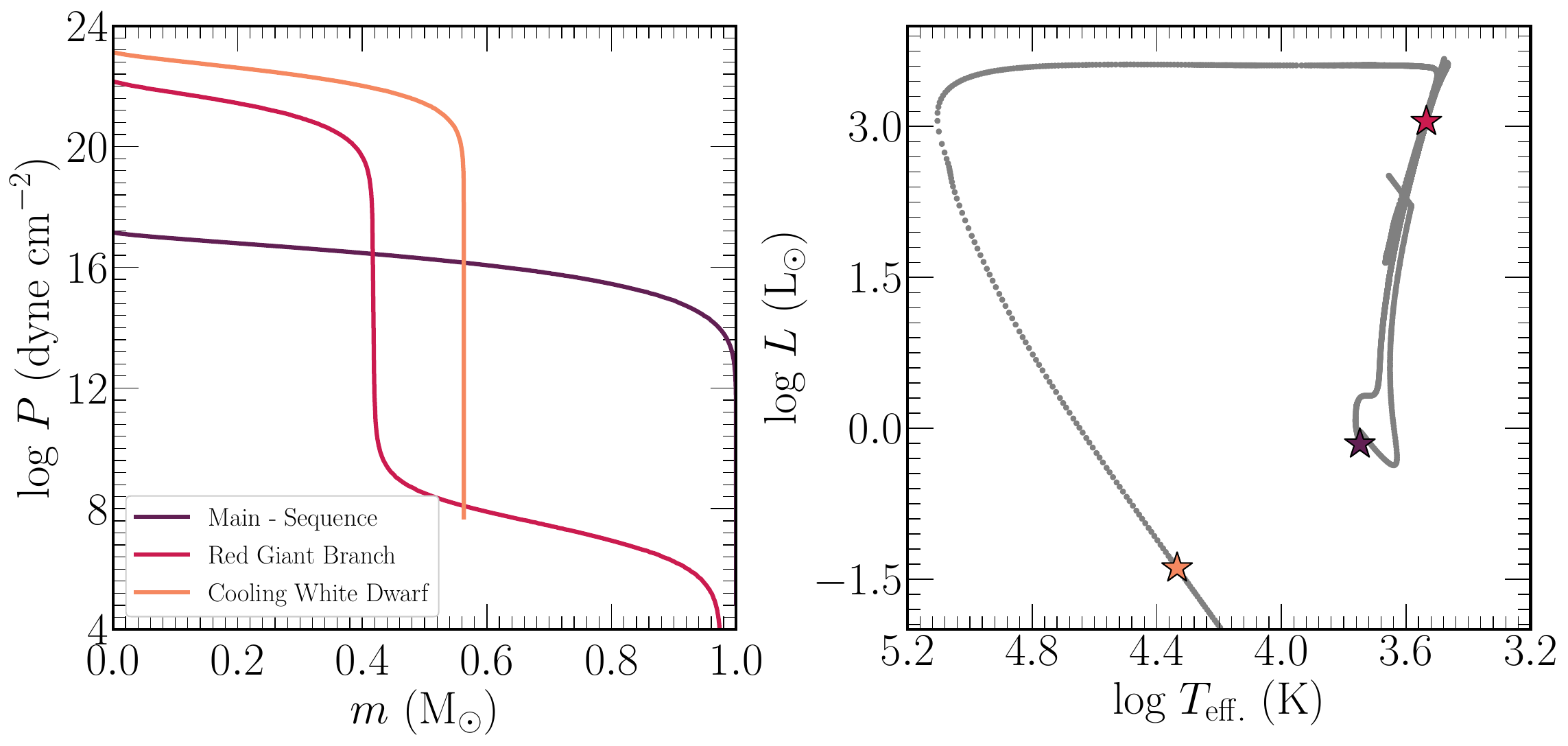}
\caption{
\texttt{MESA-Web} history and profile data for a 1 $M_{\odot}$ stellar evolutionary model at three approximate epochs: the main-sequence, red giant branch, and during the
white dwarf cooling phase.
}
\label{fig:time_evol}
\end{figure*}

\section{Learning with \texttt{MESA-Web} Output}
\label{sec:output}

As a part of the output data, learners receive an MP4 movie showing profile and history quantities of their stellar model. 
The movie is a time series of grid plots created using \texttt{MESA}'s internal \texttt{pgstar} plotting routines to 
showcase general structure and surface properties.
For example, a frame from the movie for a non-rotating 1 $M_{\odot}$ stellar model is shown in Figure~\ref{fig:grid}. 

The output also includes
numbered \texttt{profileX.data} where \texttt{X} corresponds to an integer profile number. The
profiles are produced at a learner specified interval and the \texttt{profiles.index} file is provided as a key to their 
correspondence. The profile data contains stellar structure information for 56 quantities as a function of mass coordinate. 
These data are useful for learning assignments which require information about cell-specific properties. 

The output also includes a \texttt{trimmed\_history.data} file containing the traditional "evolutionary track" data for total or cell-specific 
information at all timesteps.  For example, this file would include the surface luminosity for the stellar model as a function of 
timestep. The history file currently contains information about 57 different quantities
recorded at every timestep taken by the stellar model.  
In Figure~\ref{fig:time_evol}, we further highlight the utility of data provided by \texttt{MESA-Web} showing the 
time evolution of a 
subset of the quantities depicted in Figure~\ref{fig:grid}.  These history values and specific structure profiles are plotted at various evolutionary epochs: during the 
main-sequence, the red giant branch, and the white dwarf cooling phase. 

\section{\texttt{MESA-Web} in the classroom}
\label{sec:classroom}

Our goal is to provide open access educational materials for learning about stars. 
A step towards implementing this goal is
developing and aggregating a prototype series of lesson modules that 
address educational needs ranging from high school to graduate school.
Each module will include real-world cases and \texttt{MESA-Web}
exercises addressing a specific topic: essential principles, a
listing of \texttt{MESA-Web} settings, interpreting the results returned by \texttt{MESA-Web} in
relation to the essential principles, and providing assignments for a
learner’s current and future studies. Through hands-on exercises, learners 
gain new knowledge about stars that are not possible with traditional
textbook centric material.

Open-ended computational projects, such as those in many stellar courses, 
can have a positive impact on learning \citep{Odden:2019aa}.
The importance of computational literacy in astrophysics has also
been highlighted \citep{zingale_2016_aa}.

An anonymous survey suggests that an approximately equal number of undergraduate and graduate courses 
have utilized \texttt{MESA-Web} and that the majority of these courses are taught once per academic year. 
In 62\% of these courses, \texttt{MESA-Web} is used repeatedly during the term while the remainder (38\%) used it once.
About 55\% of the respondents said their courses had no more than 20 learners and 45\% said their courses had more than 20 learners. 
The survey also solicited suggestions for future expansion and the two most requested 
features are the ability to run giant planet models and the ability to compute 
pulsation modes with \texttt{GYRE} \citep{townsend_2013_aa,townsend_2018_aa}.  Both capabilities already exist in \texttt{MESA}. 

\subsection{Sustained collaboration and integration of \texttt{MESA-Web} in the classroom}

A goal of \texttt{MESA-Web} is to foster collaboration for educators around the globe and increase the efficacy 
of technology in astronomy courses at large. To this end, we have created a \href{https://zenodo.org/communities/mesa-web/}{Zenodo \texttt{MESA-Web} community hub} 
where educators can search or share examples of \texttt{MESA-Web} use in the classroom. 
We encourage educators to share the following
\begin{itemize}
\item Markdown (\texttt{`.md`}) file describing Lesson:
\subitem Instructor information and contact
\subitem Course description
\subitem Lesson learning objective and plan
\subitem Information regarding textbooks/materials utilized
\item Assignment(s)
\item Analysis workbooks (\texttt{Jupyter,  GoogleColab})
\item Data Analysis Scripts (Instructor Version)
\item Lecture Notes
\end{itemize}
as part of their upload. 
Our next step is to develop prototypes, leveraging a decade's worth of guiding learner experiences
during the successful MESA Summer Schools.
A MESA Summer School offers participants
a week of extensive hands-on labs to gain familiarity with \texttt{MESA}  and
learn how to use \texttt in their own education and research activities.
About 20\% of the time is spent in a lecture format;
participants spent most of their time setting up, running, and interpreting \texttt{MESA} models in small groups.
One teaching assistant at each table of three learners ensures
hands-on participation and energetic interactions among the
participants, teaching assistants, and lecturers. 
Topics vary from year-to-year, highlighting \texttt{MESA}'s flexible education and science capability, 
and were led by experts in their respective fields.
The cohort of instructors, teaching assistants, and participants 
(amateurs, undergraduates, graduates, postdocs, and faculty) now number over 500. They are
creating their own \texttt{MESA} infrastructures at $\simeq$100 institutions
around the world, which accelerates education and discovery. \texttt{MESA} Summer School 
material from the past decade is stored on 
\mbox{\href{https://zenodo.org/communities/mesa/}{Zenodo}}
and aggregated chronologically at 
\mbox{\href{https://cococubed.com/mesa_market/education.html}{this URL}}.

\begin{figure}[!t]
\centering
\includegraphics[width=\columnwidth]{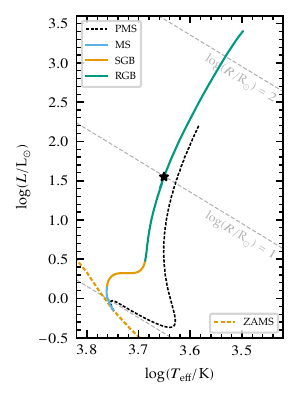}
\caption{
\texttt{MESA-Web} evolutionary track in the Hertzsprung-Russell diagram for a
for a $1\,M_{\odot}$ \texttt{MESA-Web} model, spanning the pre-main sequence (black dashed), 
main sequence (blue), sub giant branch (gold), and red giant branch (RGB; green) phases
The asterisk marks the case plotted in Figure \ref{fig:fig23p2}, and the gold dashed line marks the zero-age main sequence.
}
\label{fig:fig23p1}
\end{figure}

\begin{figure}[!t]
\centering
\includegraphics[width=\columnwidth]{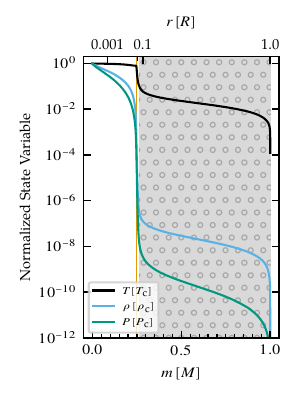}
\caption{
The state variables temperature $T$, density $\rho$ and pressure $P$ (in units of their central values 
$T_{\rm c}$\,=\,3.65$\times$ 10$^7$ K, 
$\rho_{\rm c}$\,=\,1.90$\times$10$^5$ g cm$^{-2}$,
$P_{\rm c}$\,=\,2.00$\times$10$^{21}$ Ba), 
plotted as a function of interior mass $m$ for the 1\,$M_{\odot}$ model at 
$R$\,=\,10\,$R_{\odot}$ (marked by the asterisk in Figure \ref{fig:fig23p1}). 
The ticks at the top mark the position of layers with radial coordinates 
$r$\,=\, 0.001, 0.01, 0.1 and 1 $R$. The gold-shaded region indicates 
the (very narrow) hydrogen-burning shell; 
the gray-shaded/dotted region indicates the extended convective envelope.
}
\label{fig:fig23p2}
\end{figure}


\subsection{Example: The Red Giant Branch}
The following example is part of the instructor submitted collection of \texttt{MESA-Web} learning 
materials available publicly at our \href{https://zenodo.org/communities/mesa-web/}{Zenodo \texttt{MESA-Web} community hub}.

\subsubsection{Learning Objective}

\begin{itemize}
\item[$\bullet$] Explain what happens when a Sun-like star runs out of hydrogen fuel in its core.
\end{itemize}

\subsubsection{Concept: Evolution onto the Red Giant Branch}

When a star runs out of hydrogen at its center, its time on the main
sequence has reached an end, and it embarks on a series of dramatic
changes. For low- and intermediate-mass stars like the Sun, hydrogen
burning continues in a shell around the core (now composed
predominantly of helium), and the star evolves to the right in the
Hertzsprung-Russell diagram. This phase is known as the sub giant
branch (see Figure \ref{fig:fig23p1}). 

Eventually, the star becomes mostly convective and the curve transitions to
more-vertical evolution in the HR diagram. This phase is known as the
red giant branch (RGB), because the star becomes very large
($R$\,$\ge$\,100$R_{\odot}$) and appears red due to its low effective
temperature.

\subsubsection{Concept: Core-Envelope Dichotomy on the RGB}

A characteristic property of stars on the RGB is the division of the star into two dichotomous regions.
One, a high-density radiative core composed primarily of helium (plus a small amount of metals). 
The core can encompass a significant fraction of the star's mass, but spans only a small fraction of the star's radius.
Two, a surrounding low-density convective envelope composed of hydrogen-rich material. 
The envelope contains the remainder of the star's mass, and spans almost all of the star's radius.

Figure \ref{fig:fig23p2} illustrates the marked contrast between the core and the envelope, 
for a 1\,$M_{\odot}$ \texttt{MESA-Web} model at $R$=10$R_{\odot}$ on the RGB. 
Note how the density at the top of the core is 5 orders of magnitude larger 
than the density at the bottom of the envelope. For this \texttt{MESA-Web} model, 
the core contains around 25\% of the star's mass, but spans only about 0.3\% of its radius. 
The envelope encompasses almost all of the remaining mass and radius. 
Because 95\% by radius of the star is convective, its evolution in the Hertzsprung-Russell diagram (Figure \ref{fig:fig23p2}) 
has a steeper slope.

\subsubsection{Exercises:}

\begin{itemize}
\item[$\bullet$] Enter the inlist of Figure \ref{fig:submission} (all default values) on \texttt{MESA-Web}.
Click submit. Explore the returned movies, plots, and plain text data files from which additional 
analysis can be performed. Explore the documentation of the \href{http://www.astro.wisc.edu/~townsend/static.php?ref=mesa-web-output}{\texttt{MESA-Web} output}.

\item[$\bullet$] Using the provided \texttt{Jupyter} notebook, reproduce Figures \ref{fig:fig23p1} and \ref{fig:fig23p2}.

\item[$\bullet$] Make drawings and annotations such that you can use 
it to be able to tell someone else how a star goes from the Main Sequence to the Red Giant Branch stage, 
describing the logic of how  the core contracts/expands,
how the star moves in the HR diagram, 
the means of energy transport, and 
the relevant nuclear reactions. 
Bring your drawings to class and use it (and nothing else) to tell the story of this 
phase of stellar evolution to another learner -- then exchange your roles.

\item[$\bullet$] The history file, named \texttt{trimmed\_history.data}, provides general information 
about the entire stellar model as a function of time. The file consists of a few header lines 
giving global data, followed by a sequence of rows correspond to individual timesteps. 
The columns of each row contain physical characteristics of the model. 
In a plot, compare the dynamical, Kelvin-Helmholtz, and nuclear timescales as a function of time
in the history file with analytical approximations for these timescales
as the stellar model goes from the Main Sequence to the Red Giant Branch stage.
Additional columns in the history file may be useful for evaluating the analytical approximations.

\end{itemize}

\begin{figure}[!htb]
\centering
\includegraphics[width=\columnwidth]{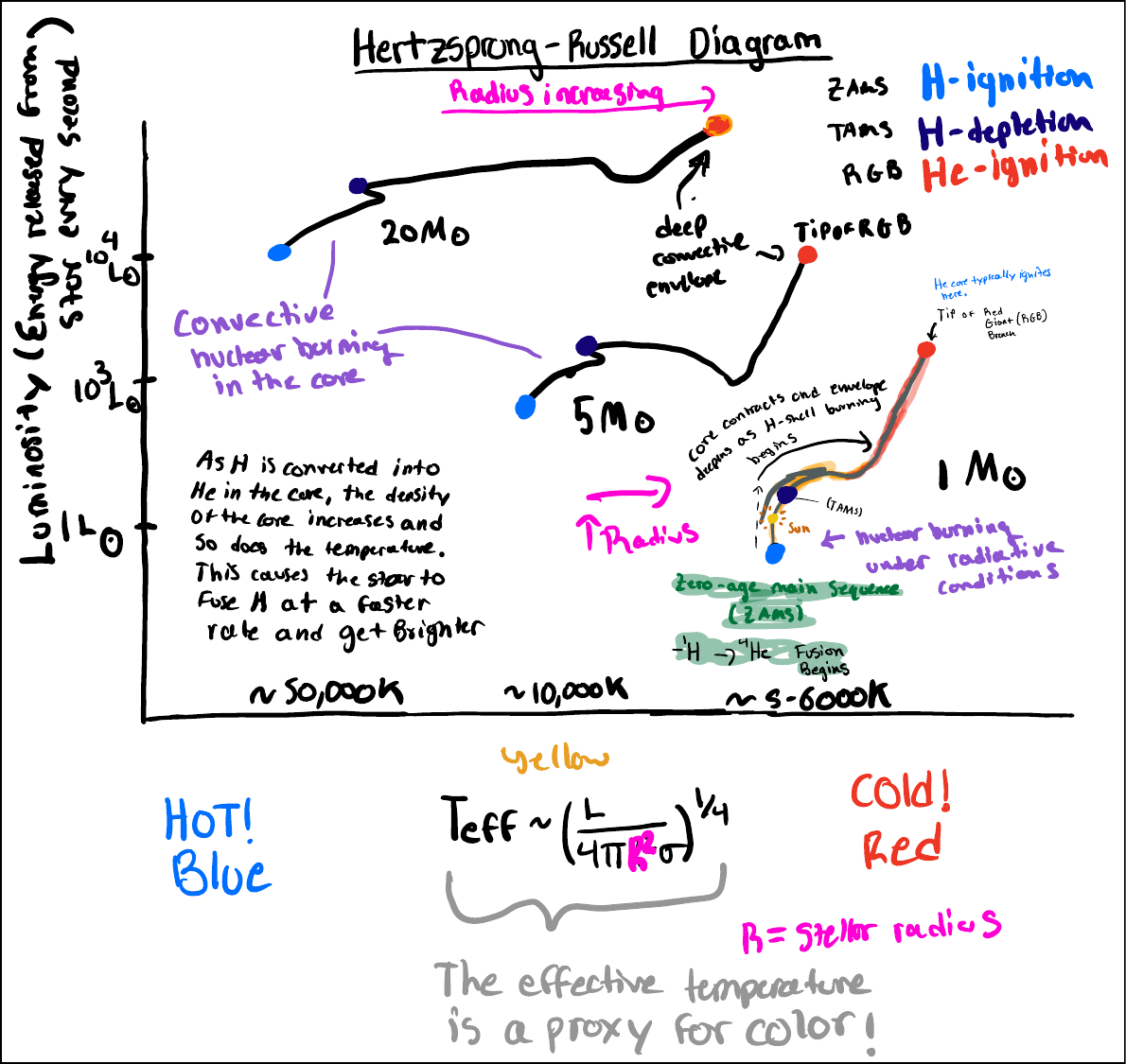} 
\hbox{ }
\includegraphics[width=\columnwidth]{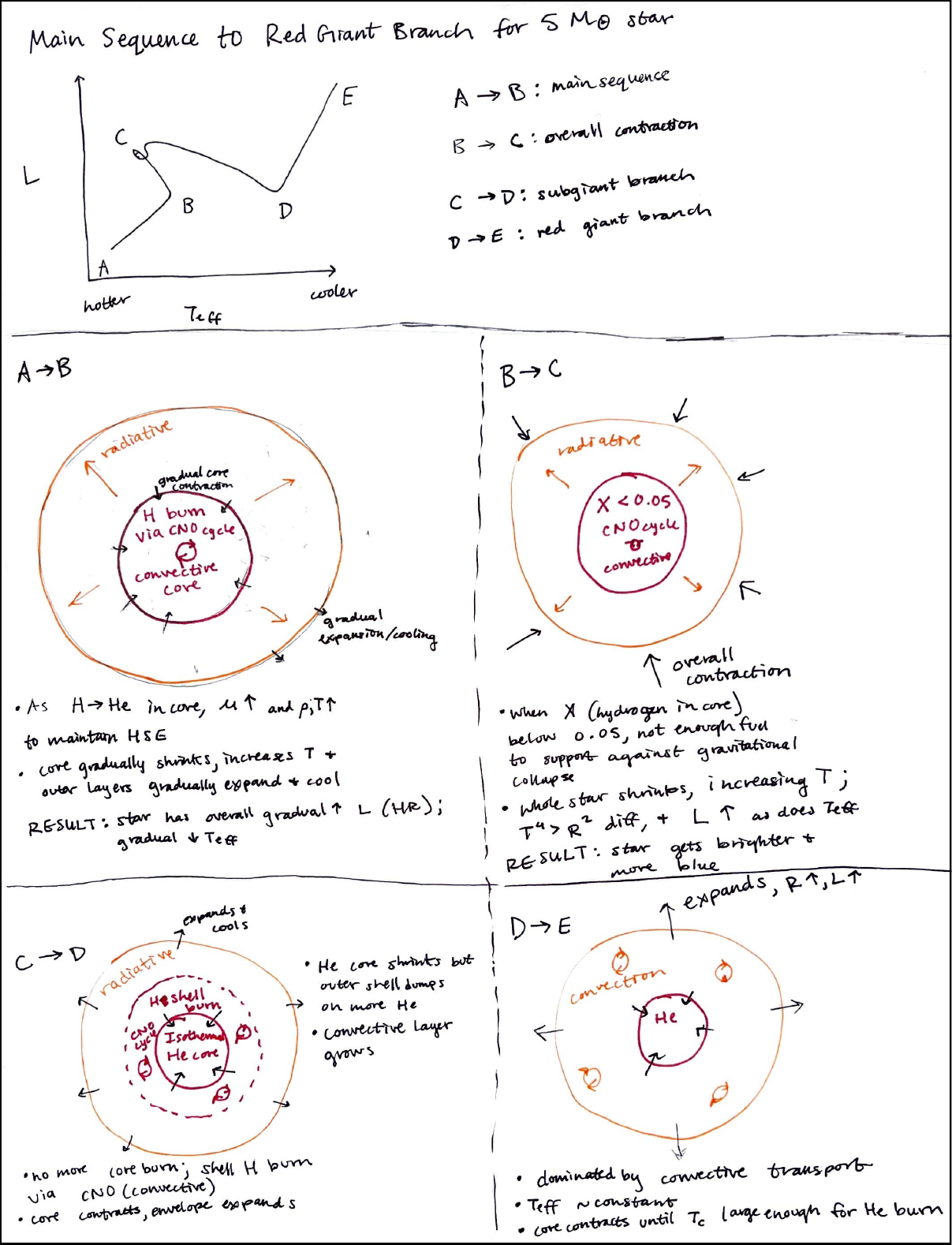}
\caption{
Two learner's responses to the drawing exercise in The Red Giant Branch Example.
}
\label{fig:ebf}
\end{figure}

Figure \mbox{\ref{fig:ebf}} shows two learner's submissions in response to the third bulleted homework exercise suggested above. 
The first example was created with a digital illustration software instrument. 
Highlights of this assignment submission include showing stellar evolution tracks for three different initial masses,
identifying the means of energy transport in the core region,
defining the axis labels,
and marking the starting and ending locations with fuel ignition events.
The second example was created with colored pens on paper, which was subsequently digitally imaged.
Highlights of this assignment submission include showing an analysis of changes in the thermodynamics and stellar structure due to nuclear burning, 
distingushing the global means of energy transport, and 
identifying astronomy names for groups of stars at different points along the evolution's trajectory.
Both submissions were adjudicated to have met the learning objective listed in this Red Giant Branch Example.

\subsection{Example: Stellar Structure and Evolution}

The following describes an example that is part of the instructor submitted collection of \texttt{MESA-Web} learning 
materials available publicly at our \href{https://zenodo.org/record/6798742#.Ysx0_-zMLWw}{Zenodo \texttt{MESA-Web} community hub}.
The lesson materials are part of a Stony Brook University's AST 341 \emph{Stars and Radiation} course.


Stony Brook's AST 341 \emph{Stars and Radiation} class is an upper-level Astronomy class for majors.  Until recently,  students were not required 
to have any computing prerequisite,  so having the students install and run \texttt{MESA} is not appropriate for the class.  Instead, they were given \texttt{MESA-Web} 
models to explore (the instructor generated a few,  just to ensure that there was some data without overloading \texttt{MESA-Web}, but the students were 
also encouraged to run their own).  Since the students may not have python experience, the \texttt{MESA} output was converted to columnar data with 
\texttt{py\_mesa\_reader} (scripts were available for students to do this on their own as well).   The assignment had them verify some of the trends that 
we explored in class in our discussion of stellar structure and evolution, and referred to the discussion in the class text,  Prialnik's \emph{An Introduction 
to the Theory of Stellar Structure and Evolution} \citep{prialnik_2009}.  For the graduate version of this class, students were given a final project where they could explore 
some aspect of stars.  Several students chose \texttt{MESA} projects, and most used \texttt{MESA-Web}, exploring, for example, the effect of metallicity on stellar 
structure, the dependence of WD mass on initial progenitor or wind model.

\section{Summary}
We have presented \href{http://user.astro.wisc.edu/~townsend/static.php?ref=mesa-web-submit}{\texttt{MESA-Web}}, a cloud resource with an online interface to the Modules for Experiments in Stellar Astrophysics (\texttt{MESA}) software instrument. The goal of \texttt{MESA-Web} is to remove barriers to access and to provide open educational resources for learning about the evolution of stars. Since its inception in 2015, \texttt{MESA-Web} has evolved over 10,000 models to over 2,200 unique learners and currently performs about 11 jobs per day. \texttt{MESA-Web} has had several major improvements in capabilities and available resources over the years, making it a powerful tool for education and scientific investigations.

To help facilitate a community of educators and access to materials that have leveraged \texttt{MESA-Web} in the classroom, we created a \href{https://zenodo.org/communities/mesa-web/}{Zenodo \texttt{MESA-Web} community hub}. This community hub serves as a central repository for sharing materials and fostering collaboration among the \texttt{MESA-Web} community. We encourage educators to help contribute to and support this hub to help aide in the continued success of \texttt{MESA-Web} as an educational resource for years to come.

\section{Declarations}

\subsection{List of abbreviations}

\begin{itemize}
\item \texttt{MESA} - Modules for Experiments in Stellar Astrophysics
\item HR - Hertzsprung-Russell
\item RGB - red giant branch
\item WD - White Dwarf
\end{itemize}

\subsection{Competing Interests}

The authors declare that they have no competing interests.

\subsection{Funding}
The \texttt{MESA} Project is supported by the National Science Foundation (NSF) under the Software Infrastructure for Sustained Innovation
program grants (ACI-1663684, ACI-1663688, ACI-1663696).
Research presented in this article was supported by the 
Laboratory Directed Research and Development program of Los Alamos National Laboratory under 
project number 20210808PRD1.  The work at Stony Brook was supported by DOE/Office of Nuclear Physics grant DE-FG02-87ER40317.

\subsection{Author's Contributions}

All the authors contributed equally to this article.

\section{Acknowledgements}
We thank the anonymous referee for suggestions that improved this article.
We also thank the \texttt{MESA} development team for their engagment with the \texttt{MESA} Project,
and the participants of the \texttt{MESA} Summer Schools for their willingness 
to experiment with new capabilities and modalities of delivery  that influenced the development of \texttt{MESA-Web}. 
Finally, we thank Ebraheem Farag and Morgan Chidester for permission to share their homework assignment submissions in Figure \mbox{\ref{fig:ebf}}.

\bibliography{refs}

\end{document}